# The electron transfer dynamics in the contact electrification and its effects on the intensity of triboluminescence

Na Li[1], Liran Ma[2], Xuefeng Xu*[1] and Jianbin Luo[2]

[1] *School of Technology, Beijing Forestry University, Beijing 100083, China*

[2] *State Key Laboratory of Tribology, Tsinghua University, Beijing 100084, China*

**Abstract**

With the growing threat of energy crisis and the increasing need to power microelectronic devices, people are seeking potential alternative energies that can replace the conventional sources such as fossil fuels. Due to its simple structure, low cost, and high performance, the triboelectric nanogenerator (TENG) which is based on the contact electrification has become one of the most promising candidates. Although plays a crucial role in determining the amount of charge transfer of TENG, the fundamental mechanism that underlies the charge transfer in the contact electrification of insulator-insulator contacts has still not been completely understood. In this paper, the dynamics of charge transfer in the contact electrification is investigated by observing the triboluminescence of the sliding contacts and a theoretical model is proposed for the electrification of insulator-insulator contacts. In the present "Capacitor model", the electron tunneling is assumed to be the dominating mechanism in the electrification and the driving force for the electron transfer is the contact potential difference, and thus, the electrification is considered as a charging process of the capacitor formed by the contact surfaces. The consistency between the predictions by the present model and the experiments may confirm again that the electron transfer is the dominant process in the contact electrification of insulator-insulator contacts.

---

* Corresponding author. E-mail: xuxuefeng@bjfu.edu.cn

# 1. Introduction

When two materials are brought into contact and then separated, electrical charges are often generated on the contacted surfaces [1-3]. This phenomenon is widely known as contact electrification which may be our earliest knowledge about electricity. Contact electrification is ubiquitous in the daily life and will lead to detrimental consequences [4-7]. For example, in powder-handling operations, the particles charged by contact electrification can adhere onto the walls and make operations less efficient [4]. In addition, the build-up of electrical charges on the surfaces by contact electrification can result in electrostatic discharge (ESD), which can destroy electronic circuits and can lead to fire or even explosion, costing the industry billions of dollars per year [5,6].

On the other hand, the charges produced in contact electrification can be utilized in a wide range of industrial and scientific applications such as electrophotography [8] and electrostatic separation [9,10]. Based on contact electrification and electrostatic induction, a new energy harvester, i.e., the triboelectric nanogenerators (TENG), has been recently developed which can efficiently convert different types of mechanical energy, ranging from human activities to ocean waves, into electricity [11]. Due to its high-level electric output, TENG can supply power for micro-scale electronics and self-powered sensors and thus has received global attention in recent years [12-16].

Contact electrification is essentially the result of charge transfer between contacting surfaces and electrons are often believed to be the main mechanisms [1,4,12-14,17-23]. When two metals are brought together, to obtain the thermodynamic equilibrium, their Fermi levels should coincide with each other. For this purpose, electrons tend to flow from the metal with a lower work function to the metal with a higher work function until a potential difference between the contacting surfaces is established to stop the electron transfer [1,4,17,19,20]. The work function of a metal is defined as the energy required to remove an electron from the metal surface and equals to the energy difference between the Fermi level of the metal and the vacuum level. Experiments on the electrification of metal-metal contacts indicated that the amount of the transferred electrons is proportional to the contact potential difference (CPD) which is defined as the difference in the work functions of the metals [19].

For insulators, the charges acquired after contacted with metals were found to have linear dependence on the work function of the metals [1,4,21-24]. This implies that electrification in metal-insulator contact is also a result of transfer of electrons, and further suggests that there should also exist some energy level of electrons inside the dielectric materials which can be considered as the "Fermi level" of the materials. By observing the change in the photoemission characteristic of polyethylene after the polymer is electrified by contact with metals with different work functions, Murata et al. [25,26] provided clear evidence of electron transfer as the mechanism of contact electrification in metal-insulator contacts. Further, by using triboelectric nanogenerators, Xu et al. [12] directly measured the real-time charge transfer in contact electrification as a function of temperature, and confirmed again that the electron transfer is the dominating mechanism for the metal-insulator electrification.

By measuring the contact charging of insulators when contacted by metals with known work function, the "Fermi level" of the insulators can be determined and the "effective work function" of the insulators has been obtained [1,21-24]. Further experimental evidence shows that the electrification between insulators can be predicted from knowledge of the electrification of each of the insulators by metals [1,4,27,28]. This suggests that, similar to the metal-metal contacts or the metal-insulator contacts, the contact electrification in the insulator-insulator contacts may also be dominated by the electron transfer. Assuming that electron transfer is the cause of contact electrification, Shen et al. [29] calculated the charge transfer between sapphire and quartz surfaces using the first-principles calculation. Both the direction and the magnitude of charge transfer predicted by the calculations are found consistent with the experimentally obtained results which shows that sapphire charges positively and quartz charges negatively. From electrochemical (redox) experiments, Liu [30] directly identified the electrostatic charges on Teflon produced by rubbing with Lucite as electrons rather than ions.

No matter to improve the performance of TENG or to reduce the risk of fire or explosion, the charge transfer dynamics in the contact electrification of insulator-insulator contacts plays an important role but remains elusive [1,17,18,21,22]. Measuring directly the time dependence of the charge transferred between the contacting surfaces is not easy because it is difficult to make and break contact in a short time without introducing any uncertainty. As a type of contact electrification, triboelectrification (also known as triboelectric charging) [1,21,22], which denotes the charge transfer between two surfaces in the frictional contacts, can provide a feasible method for investigating the dynamics of charge transfer in the electrification. In this paper, we focus our attention on the light emission (i.e., triboluminescence) [31-36] in the slide contacts, which arises from the dielectric breakdown of the surrounding gas in the contact gap and therefore can be used as an indicator for the charge transfer process in the contact electrification. To reveal the charging characteristics, the effects of the sliding speed on the light intensity in the sliding friction were investigated. Then a theoretical model for the charge transfer dynamics in the contact electrification of insulator-insulator contacts was proposed. Further, the present theoretical predictions were compared and found reasonably consistent with the experimental measurements, indicating that electrons tunneling may be the dominant mechanism.

## 2. Experimental section

The sliding friction experiments were conducted by using a disk-on-disk apparatus settled in a dark chamber (see Fig. 1). In the apparatus, an upper stationary disk is pressed against a lower rotating disk under a normal force $F_N$. During the sliding contacts, photons are emitted from the contact region and then are transferred via a fiber to a spectrometer (HORIBA Scientific, iHR320). The spectrometer is equipped with a photomultiplier tube (PMT, Hamamatsu R928) which is sensitive to wavelengths from 200 to 900 nm. Both the spectrum and the intensity of the light emission are recorded and analyzed by a computer connected to the spectrometer. Compared with the electrical measurements, photon counting during the electrification process has the advantages of accuracy and stability.

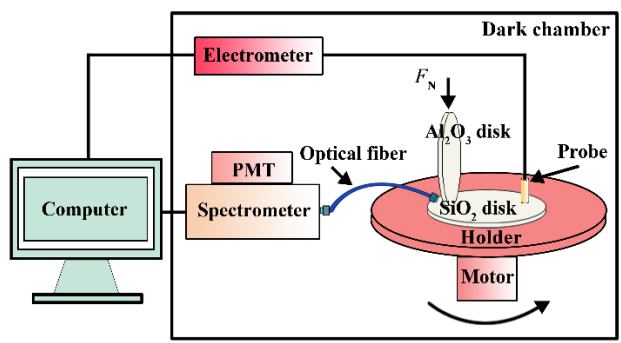

**Fig. 1.** The apparatus used to measure the photons emitted from the sliding contact and the charge density on the frictional surface.

The samples used here are round disks of $Al_2O_3$ crystal, $Al_2O_3$ ceramic, $SiO_2$ crystal and $SiO_2$ glass with a diameter of 30 mm and a thickness of 2 mm (Shanghai Daheng). The measurements were carried out in the ambient air with a temperature of about 20°C and a relative humidity about 22%. In the experiments, $Al_2O_3$ disks are always used as the upper disks and $SiO_2$ disks the lower ones. The diameter of the sliding track on the disk surface is about 18 mm and the down force is fixed ($F_N$=0.89N).

## 3. Results and discussion

### 3.1 Relationship between triboluminescence and electrification

To clarify the origin of the triboluminescence in the present experiments, spectrums of the light emission during the sliding friction under different experimental conditions were measured. In the spectrums, sharp peaks in the region of ultraviolet (UV), visible, and infrared (IR) have been observed (see Fig. 2). Comparison between the spectrums indicates that the peaks in all the experiments almost completely coincide, although the intensity of the peaks may be different for each measurement. This implies that the light emission in all the present experiments has the same origin.

By comparing the peaks in the spectrums of triboluminescence with those from the gas discharge, the light emission in the present experiments can be attributed to the discharge of the gas in the contact gap, which is in turn a result of the contact electrification during the friction process [36]. When friction, the contact electrification will generate an intense electric field in the gap of the sliding contact. Under the action of the electric field, electrical discharge of the ambient gas in the gap may occur, and as a result, photons may be emitted from the contact gap.

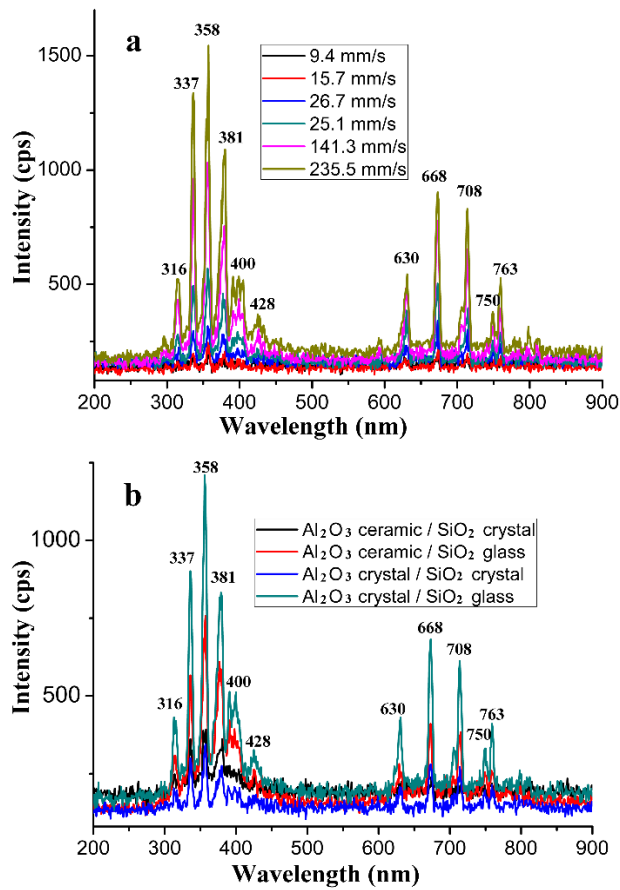

**Fig. 2. (a)** Spectrums of triboluminescence under different sliding speeds. The friction pair used is "Al$_2$O$_3$ crystal/SiO$_2$ glass". **(b)** Spectrums of triboluminescence of different friction pairs. The relative sliding speed used is 0.19 ms$^{-1}$.

*3.2 Dependence of light intensity on sliding speed*

To reveal the charge transfer dynamics in the slide contacts, investigating the dependence of the luminescence intensity on the sliding speed may be useful. For such a goal, the intensities of the emitted light under different sliding speeds were measured and illustrated in Fig. 3. From the figure it can be seen that the intensity of the emitted photons increases monotonically with the relative sliding speed.

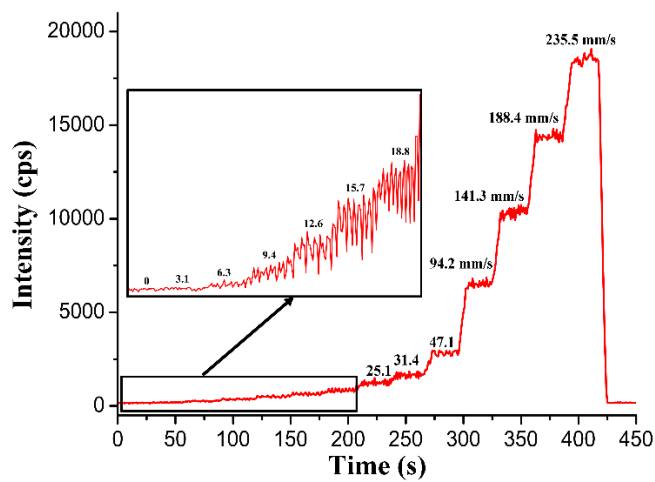

**Fig. 3.** The variation of the light intensity with time under different relative sliding speeds. The friction pair used here is $Al_2O_3$ crystal / $SiO_2$ glass. The inset in the figure shows the enlargement of the lower speed region.

To perform quantitative analysis on the speed dependence of the light intensity, the measured intensities of each speed were first averaged and then subtracted by the noise value. In order to focus our study merely on the effects of the sliding speed and to eliminate as much as possible the influences of other related factors such as material, temperature and humidity, the obtained light intensity of each speed was further normalized compared to the intensity value under the highest speed in the experiments. After normalization, we found that the variation curves of the light intensity with the sliding speed are almost coincident for all the friction pairs (see Fig. 4(a)). Meanwhile, the result shows that the increase of the light intensity with the sliding speed can be divided into two regions: a plateau region in which the light intensity remains almost unchanged and a rapid increase region in which the light intensity increases dramatically.

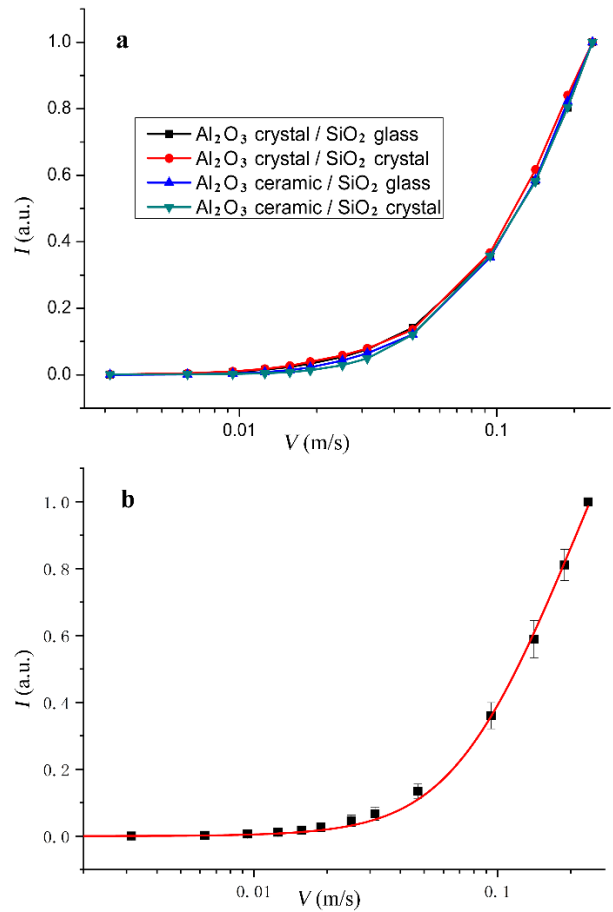

**Fig. 4.** The variation of the normalized triboluminescence intensity *I* as a function of the relative sliding speed *V*. **(a)** The data points of each friction pair are the average of 9 sets of experimental data. **(b)** The data points represent the mean values of 36 sets of experimental data including that of all the four friction pairs and the error bars represent the standard deviation. The solid line is the best fit to Equation 2 and yields $I = 50.562V^2(1-e^{-0.21183/V})^2$, where *V* is given in ms$^{-1}$. The coefficient of determination of the fit is 0.99876.

*3.3 A theoretical model for charge transfer dynamics in contact electrification*

To explain the variation of the light intensity with the sliding speed, a "Capacitor model" was proposed for the charge transfer dynamics in the contact electrification. The model assumes that, in the electrification of insulator-insulator contacts, the species that carries the charge from one surface to another are electrons and the driving force is the contact potential difference between the contacting surfaces. Once two surfaces are brought together, the contact potential difference can drive the electrons to tunnel from the surface with a lower work function to the surface with a higher work function [1,4,17-23]. If the contact potential difference $V_{CPD} = |\varphi_2 - \varphi_1| \ll \bar{\varphi}$, where $\varphi_1$ and $\varphi_2$ are the work functions of the two surfaces respectively and $\bar{\varphi} = (\varphi_1 + \varphi_2)/2$ is the mean barrier height through which electrons penetrate, the tunnel current during the electrification process will be a linear function of the potential difference between the surfaces [37]. This Ohmic characteristic implies that the electron transfer by tunneling between the contacting surfaces in the electrification can be considered as a charging process of the capacitor formed by the contact region of the two surfaces (see Fig. 5).

The effective work functions of $SiO_2$ and $Al_2O_3$ are 5.5eV and 5.25 eV respectively [23], and thus electrons will moves from $Al_2O_3$ disks (upper disks) to $SiO_2$ disks (lower disks) when they are brought into contact. According to the capacitor model, in each rotation of the lower disk, the charge quantity transferred to the lower surface per unit circumferential length of the contact circle can be expressed as

$$Q = Q_0 \left(1 - e^{-t_C/\tau}\right) = Q_0 \left(1 - e^{-L/V\tau}\right) \tag{1}$$

where $Q_0$ is charge quantity transferred at $t_C = \infty$, $t_C = L/V$ is the time period for which a given point in the contact track of the lower surface has been in contact with the upper disk in each revolution of the lower disk, $L$ is the width of the contact area, $V$ is relative sliding speed, $\tau = RC$ is the time constant of the equivalent $RC$ circuit (see Fig. 5), $R$ is the tunneling resistance, $C = \dfrac{\varepsilon_0 S}{d}$ is the capacitance of the capacitor, $\varepsilon_0 = 8.854 \times 10^{-12}$ Fm$^{-1}$ is the permittivity of free space [38], $S$ is the contact area, and $d$ is the separation between the surfaces at contact. Calculated by Hertz contact theory [39,40,41], $L$ is about 23.43 µm in the present case. The result also shows that the vertical deformation of the $SiO_2$ surface (≈0.023µm) when contact is larger than that of the $Al_2O_3$ surface (≈0.007µm) because $Al_2O_3$ is relatively harder than $SiO_2$. However, both of them are much smaller than the width of the contact area. This means that we can reasonably consider the contact surface as a flat surface to support the capacitor model (see Fig. 5).

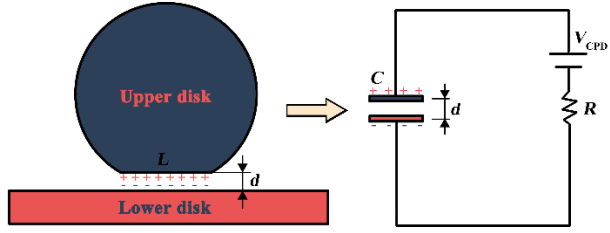

**Fig. 5.** "Capacitor model" for the contact electrification in sliding contacts. The right side of the figure shows the equivalent *RC* circuit. $V_{CPD}$ is the contact potential difference between the contacting surfaces, *R* is the tunneling resistance, *C* is the capacitance of the capacitor formed by the two contacting surfaces, and *d* is the separation between contacting surfaces at contact.

The experimental measurements show that the light intensity is nearly constant at a specific sliding speed, (see Fig. 3). This means that a balance should occur between the charging and the discharging. In such an equilibrium state, the number of electrons $N_{er}$ per unit time released from the negatively-charged surface into the contact gap should be statistically equal to the number of electrons transferred per unit time to the lower surface, i.e., $N_{er} \sim QV$. Further measurements indicate that, when the equilibrium state is reached, the lower SiO$_2$ disk surface is always charged negatively and the electric potential of the surface $V_{surf}$ is also approximately proportional to the value of $QV$ (see Fig. 6). Considering that, in the discharging process, the number of photons excited by a single electron released from the electrified surface should be proportional to the energy the electron gains from the electric field when transferring from one surface to another, i.e., be proportional to the electric potential of the lower disk surface $V_{surf}$, the intensity of the light emission *I* during the sliding process is proportional to $N_{er} V_{surf}$, and thus can be expressed as

$$I \sim N_{er}V_{surf} \sim (QV)^2 = AV^2\left(1-e^{-L/V\tau}\right)^2 \quad (2)$$

where *A* is a fitting parameter.

From the formula about the tunnel current density given by Simmons [37], the tunneling resistance between the contacting surfaces in the contact electrification can be deduced as

$$R = \frac{d}{(2m\bar{\varphi})^{1/2}(e/h)^2 S} \exp(\frac{4\pi d(2m\bar{\varphi})^{1/2}}{h}) \tag{3}$$

where $m=9.109\times10^{-31}$ kg is the electron mass [38], $e=1.602\times10^{-19}$ C is the elementary charge [38], and $h=6.626\times10^{-34}$ Js is Planck constant [38]. Substituting $\tau = RC$ and $C = \frac{\varepsilon_0 S}{d}$ in Eq. (3) gives

$$\tau = 2.804\times10^{-13} \exp(10.246\bar{\varphi}^{1/2}d)/\bar{\varphi}^{1/2} \tag{4}$$

where $d$ is in nm, $\bar{\varphi}$ in eV, and $\tau$ in ms.

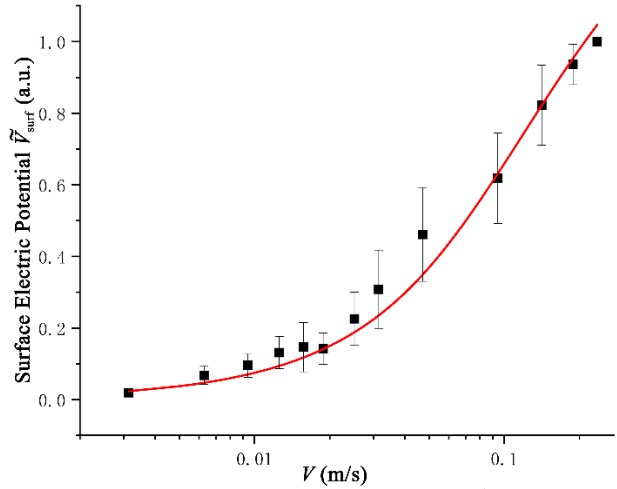

**Fig. 6.** The variation of the normalized electric potential of the lower disk surface $\tilde{V}_{surf}$ as a function of the sliding speed $V$. The solid line is a best fit and yields $\tilde{V}_{surf} = 7.4933\times V(1-e^{-0.21183/V})$, where $V$ is given in ms$^{-1}$. The coefficient of determination of the fit is 0.98268.

### *3.4 Experimental corroboration*

To validate the capacitor model mentioned above, comparisons between the theoretical predictions and the experimental measurements were performed. The fitting between the measured data and Eq. (2) shows excellent agreement (see Fig. 4). From the fitting results, the charging time constant of the contact electrification was calculated as: $\tau= 0.111$ ms. For the insulator-insulator contacts, this time required for the charging process is relatively short considering that the large forbidden energy gap in the insulators will result in a longer time needed for achieving the thermodynamic equilibrium between electrons of the contacting insulators.

This inconsistency in the charging time can be explained by the "surface state" of electrons in the dielectric materials. Because the environment of atoms on the solid surface differs from that of the bulk ones, the surface atoms might be expected to have electron states (i.e., surface states) whose energy is within the energy gap of the materials [1,4,17,18,22,42]. When an insulator is contacted with another, it is the electrons in the surface states of the insulators will be in equilibrium with each other. Driven by the difference in the effective work functions of the two insulators, electrons in the surface states may transfer almost instantaneously from the surface with a lower work function to the surface with a higher work function [1,21,22]. For insulator-insulator contacts, the surface state theory was found in good agreement with the experimental measurements [4,17,18,43].

Substituting $\varphi_1 = 5.5$ eV for $SiO_2$, $\varphi_2 = 5.25$ eV for $Al_2O_3$ [23], and $\tau = 0.111$ ms in Eq. (4) gives $d=1.16$ nm. The value of $d$ obtained here is larger than the separation between smooth surfaces at contact (≈0.4 nm [44,45]). The reason is that the present model assumes perfectly smooth surfaces in the contact region but in reality the surfaces are always to some extent rough. Because the maximum tunneling distance of electrons between contacting surfaces was approximately 2 nm [1,46,47], the value of $d$ calculated by the present model from the experimental measurements on rough surfaces should be larger than the separation between smooth surfaces at contact. Although the exact solution of the "effective separation" between two rough surfaces in the present model is difficult to be calculated, it is reasonable to conjecture that the "effective separation" should be between 0.4 nm and 2 nm. Taking both the surface roughness and the maximum tunneling distance into account, the present value of $d$ calculated by the "Capacitor model" may be reasonable.

Besides the electron transfer, the contact electrification between insulators can also be attributed to the transfer of ions or materials [1,2,4,17,18]. But, a great deal of experimental evidence shows that the electrification of $Al_2O_3$-$SiO_2$ contact can be mainly attributed to the electron transfer [12,23,29]. By using a Ti-Al$_2$O$_3$ TENG and a Ti-SiO$_2$ TENG, the real-time charge transfer in contact electrification as a function of temperature have been directly measured. The results show that the electron transfer is the dominating mechanism for the electrification of Ti-Al$_2$O$_3$ contact and Ti-SiO$_2$ contact. This may imply that the electrification of Al$_2$O$_3$-SiO$_2$ contact should also be dominated by the electron transfer [12]. The consistency between the first-principles calculations and the experiments confirms that electron transfer is the cause of contact electrification in Al$_2$O$_3$-SiO$_2$ contact [29]. By measuring the contact charging when contacted by metals with known work function, the effective work functions of Al$_2$O$_3$ and SiO$_2$ were obtained [23].

Meanwhile, although the overall electrical effect of the electron transfer cannot be distinguished from that of the ion transfer, the charging dynamics of the contact electrification induced by different charged species should be quite different because a proton has a mass which is about 1836 times that of an electron and other ions are even heavier, and thus positive and negative ions are not likely to tunnel between surfaces. Therefore, the consistency between the theoretical predictions and the experimental measurements in the present work may confirm again that the electron transfer is the dominant cause of contact electrification between insulators.

## 4. Conclusions

In the present work, the charging dynamics in the electrification of insulator-insulator contacts has been studied by performing measurements on the triboluminescence from frictional contacts. The measured light spectrums indicates that the triboluminescence in the slide contacts is the result of the gas discharge in the contact gap and therefore can be used to characterize the charging process of the contact electrification. Further experiments show that the light intensity increases very slowly initially, and then grows sharply as the sliding speed increases. To explain such a variation of the light intensity with the speed, a theoretical model for the charge transfer dynamics in the contact electrification of insulator-insulator contacts has been proposed, assuming that the electron tunneling between the contacting surfaces is the dominant mechanism. In the model, the electrification is considered as a charging process of the capacitor formed by the contacting surfaces and the driving force is the contact potential difference. The theoretical predictions of the present model have been compared with the experimental measurements and were found in reasonable agreement.

Because of the assumption of perfectly smooth surfaces in the contact region, the theoretical model presented here is not appropriate for the rough contacting surfaces. But, in spite of its simplifications and limitations, the present model reflects the essential features of charge transfer dynamics in the contact electrification and can be used as a framework for addressing more complicated situations in which the morphology of the contacting surfaces should be considered. Our measurements support in principle the "Capacitor model" and show that the amount of charge transfer during contact electrification can be effectively predicted and controlled. The present work may provide theoretical basis for a wide range of applications involving contact electrification such as the optimized design of TENG.

## Acknowledgements

This research was supported by the National Natural Science Foundations of China (51575054 and 51527901).